\documentclass[a4paper,11pt]{article}

\usepackage{textcomp}
\usepackage{amssymb}
\usepackage{units}

\usepackage[overload]{textcase}

\usepackage{graphicx}
\usepackage{subfigure}

\usepackage[labelsep=period]{caption}

\usepackage{amssymb}

\usepackage{amsmath}
\usepackage{mathabx}

\usepackage[sort&compress]{natbib}
\bibpunct{[}{]}{,}{n}{}{}

\begin{document}

\thispagestyle{plain}
\setcounter{page}{1}

\vfill

\begin{center}
 {\large Industrial Research Limited Report No. 2385, December 2009}\\
   \vspace{0.3cm}
   {\Large \textbf{Detection of polystyrene sphere translocations using resizable elastomeric nanopores}} \\
   \vspace{0.3cm}
   {\large Geoff R. Willmott$^\dag$ and Lara H. Bauerfeind}\\
   {\large \emph{Industrial Research Limited, 69 Gracefield Rd, PO Box 31-310, Lower Hutt 5040, New Zealand}}\\
	\vspace{0.3cm}
	{\large $^\dag$Corresponding author}\\
	\vspace{0.3cm}
	{\large Email: g.willmott@irl.cri.nz}\\
	\vspace{0.3cm}
    {\large Phone: (64) (0)4 931 3220}\\
    \vspace{0.3cm}
    {\large Fax: (64) (0)4 931 3117}\\

\renewcommand{\abstractname}{}

\begin{abstract}
\noindent Resizable elastomeric nanopores have been used to measure pulses of ionic current caused by carboxylated polystyrene spheres of diameter 200~nm and 800~nm. The nanopores represent a novel technology which enables nanoscale resizing of a pore by macroscopic actuation of an elastomeric membrane. Three different pores were employed with variable applied strain, transmembrane potential, particle concentration and sphere radius. Theory describing current pulse magnitude has been extended to conical pore geometry. A consistent method for interpretation of data close to the noise threshold has been introduced, and experimental data has been used to compare several methods for efficient, non-destructive calculation of pore dimensions. The most effective models emphasize the absolute pulse size, which is predominantly determined by the opening radius at the narrowest part of the roughly conical pores, rather than the profile along the entire pore length. Experiments were carried out in a regime for which both electro-osmotic and electrophoretic transport are significant.   
\end{abstract}
\end{center}


\vspace{0.3cm}
\clearpage

\section{Introduction}

Resizable elastomeric nanopores represent one of the most interesting new technologies in the burgeoning field of nanopore science. Individual nanopores in thin membranes are attracting interest due to their application to sensing of single molecules or small particles, their simplicity, and the growing capability in related fabrication and characterisation techniques. Resizable nanopores are each fabricated in an elastomeric membrane, so that the nanoscale dimensions of the pore can be altered by stretching and relaxing the membrane on macroscopic scales. An initial study demonstrated that individual double-stranded DNA molecules could be gated using this mechanism \cite{454}. More recent work has included systematic ionic current measurements and analysis of pore actuation \cite{660,771}, efforts to characterize the pores using AFM and SEM \cite{660,787}, and an initial compilation of theory, experiments and ideas relating to resizable nanopores \cite{738}. Resizable nanopore technology is being uniquely developed by Izon Science (Christchurch, New Zealand), who supplied specimens, analysis software and the qNano\texttrademark\space actuation platform for the present work. 

The fundamental, nanoscale functionality of resizable nanopores lends itself to a wide range of potential applications. Entirely new processes could arise from this concept, such as mechanical trapping of small particles or molecules, controlled mechanical tuning at the nanoscale, or localisation and confinement of reaction chemistry. However, the most immediate interest has been generated by the new functionality added to applications which have been studied using conventional (static) pores, such as translocation of particles \cite{562,530,505,764,776}. Work towards fast genomic sequencing, perhaps the most exciting potential application of nanopores \cite{566,570,599,770,678,774}, has arisen from translocation experiments using nucleic acids \cite{611}. With a resizable pore, translocation rates can be controlled, particles can be gated, and the physics of translocation can be studied in novel ways. For translocation measurements, nanopores are filled and surrounded by an aqueous electrolyte, allowing measurement of electric current when a potential is applied across the membrane. 

Resizable nanopores have further practical advantages over static pores, which have employed a range of membrane materials and nanopore fabrication methods \cite{562,505,530,772,773}. The most widely-used conventional pores are `solid-state' nanopores fabricated in thin, rigid, silicon-based \cite{562} or polymer \cite{505, 530} membranes and `biological' pores \cite{528}, especially the $\alpha$-haemolysin pore \cite{589} supported by a phospholipid membrane \cite{562,770,761,678}. In comparison with such pores, resizable nanopores are less likely to irreversibly clog, as the elastomer can be stretched to allow the pore to clear. The elastomer is light and the effective pore size for a single specimen can be varied over approximately one order of magnitude \cite{660}, enabling it to perform the tasks of several static pores. The fabrication method is quick, simple and relatively inexpensive \cite{454,738}. Elastomeric pores are chemically, mechanically and thermally stable in comparison with biological pores, which must be carefully supported under laboratory conditions. The major present drawback of solid state pores, including resizable nanopores, is poor reproducibility of geometry below $\sim$50~nm diameter \cite{767}. It is hoped that studies such as the present one will pave the way for the manufacture, characterization and widespread use of reliable, molecular-scale resizable pores.

In this article, we present experiments in which polystyrene nanospheres have been detected using resizable nanopores. The work demonstrates that this technology can be used for translocation experiments, in a manner similar to static nanopores. The experiments build on initial work \cite{738} by obtaining data using various nanopore specimens, applied strains, applied potentials, particle sizes and particle concentrations. The investigation is the first to consider the relationship between current signals, noise, particle size, particle type and pore geometry with relation to resizable pores. We address the fact that particle-induced current signals may not always indicate passage of the particle from one side of the membrane to the other. An important feature of this work is a direct comparison of existing theoretical models for relating the resistance pulse size to pore dimensions  \cite{760,759,764,767,768,752,762}, including a conical-pore approach developed in the theoretical background. An efficient, non-destructive model for determining pore size would be valuable for many envisaged applications of this technology. Particle transport is also specifically addressed, both in experiments and the identification of relevant aspects of the Space-Charge model (e.g. \cite{779,762}) in the theoretical background. 

Resizable nanopore specimens are currently available for sensing particles in the hundreds of nanometres range; the size of the polystyrene particles used here (200 and 800~nm diameter) indicates the size of the narrowest part of the pore. This size range is relevant to many viruses of interest in human and veterinary medicine, agriculture and environmental studies. Many of the recent developments in virus detection technology have required fluorescent molecular tagging of the virus. The most relevant previous work includes recent nanopore studies using particles of similar size and material \cite{768,769,752,753,778} as well as earlier work relating to the development of Coulter counters \cite{767,753,760,775,681}, which included detection of particles as small as 60~nm \cite{775}. The membranes used in experiments are relatively thick (200~$\mu$m when unstretched) in comparison with other studies. Pores are conical \cite{660,738,787}, and some current rectification, as studied in some depth elsewhere \cite{535,639,522,671,784,785}, has been observed previously \cite{660}. The pores used do not close entirely when the membrane is relaxed. 


\section{Theoretical Background}

In this Section, theory relating to two major topics is presented. Firstly, 
we recount descriptions of transport and apply them in the context of the present work. The approaches used have been widely applied to both ions \cite{506,778,753,670,510} and particles \cite{752,778,530,753,767,776} in experimental nanopore work elsewhere. Secondly, we present theory developed by deBlois and Bean \cite{760} and widely used in recent studies \cite{759,764,767,768,752,762} to describe the size of a resistive pulse generated by a spherical particle passing through a nanopore. This theory is reviewed, so that assumptions can be tested by the experimental work, and we explicitly extend the existing theory to conical geometry.

\subsection{Transport of Charged Particles}
\label{PFlux}

Several papers \cite{779,762,778,670,671} summarise theoretical approaches that have been used to study 
transport of ions and particles in pores, ranging from biological ion channels to micron-sized synthetic pores. We employ a simple approach \cite{786} devised for assessing the relative contribution of various transport mechanisms, and their functional dependencies, for experimental work. This approach is a simplification of the comprehensive, but computationally expensive Space-Charge model \cite{779,762}.

Transport of aqueous ions or larger charged particles is described by the Nernst-Planck equation. The particle flux $\mathbf{J}$ is given in terms of diffusive $\mathbf{J_{diff}}$, electrophoretic $\mathbf{J_{ep}}$ and convective $\mathbf{J_{conv}}$ contributions, by

\begin{equation}\label{eq:tripartite}
\begin{split}
\mathbf{J}&=\mathbf{J_{diff}}+\mathbf{J_{ep}}+\mathbf{J_{conv}}\\
&=-D\nabla C+\frac{\xi e}{k_B T}DC\mathbf{E}+C\mathbf{v}.
\end{split}
\end{equation}

\noindent Here, $D$ is the diffusion coefficient, $C$ is concentration, $k_B$ is Boltzmann's constant, $T$ is temperature, $\mathbf{E}$ is the electric field and $\mathbf{v}$ is the convective flow velocity. The electronic charge magnitude $e$ is multiplied by $\xi$ to find the total effective charge on the ion or particle; $\xi=-1$ for an unscreened electron. Alternative notations use the gas constant $R=k_B N_A$ and the Faraday constant $F=N_A e$, where $N_A$ is Avogadro's number. At 293~K, a dimensionless ratio for determining the significance of the diffusion term is \cite{786}, 

\begin{equation}\label{eq:diff}
\biggl\lvert\frac{\mathbf{J_{diff}}}{\mathbf{J_{ep}}}\biggr\rvert 
=\frac{0.025}{\xi V}.
\end{equation}

\noindent Generally, it is expected that $\lvert\xi\rvert\gtrsim 100$ for carboxylated polystyrene nanospheres of \emph{O}(100)~nm diameter, in which case 
the ratio in Eq.~\ref{eq:diff} is $<$~5\% for any potential above $\sim 0.01$~V. Therefore, diffusion should not significantly contribute to particle transport in the present work. Electrolyte concentration is the same on both sides of the pore, so ionic diffusion is also not important. 

\begin{figure}
\begin{center}
\includegraphics[width=8.5cm]{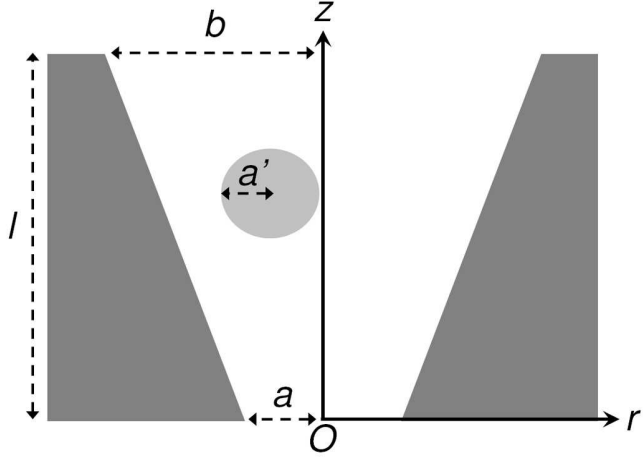}
\end{center}
\caption{A schematic cross-section through a truncated circular conical pore, showing geometrical quantities used in the theoretical analyses. Cone entry radii are $a$ and $b$ and the pore length (equal to the membrane thickness) is $l$. A translocating sphere of radius $a'$ is shown in the conical passage and a cylindrical polar co-ordinate system is defined.}
\label{FTube}     
\end{figure}

In the absence of other external forces, such as a pressure gradient, electro-osmosis is the sole mechanism for convective transport. Referring to the geometry defined in Fig.~\ref{FTube}, and using the simplified model of a long, thin cylindrical pore ($a=b=a_0$, $l\gg a_0$) to model the electro-osmotic flux ($J_z=\lvert\mathbf{J}\rvert$), we find \cite{517,786}

\begin{equation}\label{eq:vzeo}
\begin{split}
J_{z,eo}&=-\frac{\epsilon \psi_0}{4\pi\eta}CE_z\left(1-\frac{2I_1\left(\kappa a\right)}{\kappa a I_0\left(\kappa a\right)}\right)\\
&=-\frac{\epsilon \psi_0}{4\pi\eta}E_z A.
\end{split}
\end{equation}

\noindent Here, $\eta$ is the dynamic viscosity of the fluid, $\epsilon$ is the dielectric constant, $\psi_0$ is the potential of the pore wall and $\kappa^{-1}$ is the characteristic thickness of the electrical double layer at the pore wall, and $I_n$ is the $n$th-order modified Bessel function of the first kind. The subscript $eo$ refers to electro-osmotic flow. The value of $A$ increases asymptotically towards 1, and is greater than 0.9 for $a\gtrsim 20$~nm when using 0.1~M KCl ($\kappa^{-1} = 9.62$~x~$10^{-8}$~m$^{-1}$ \cite{564}). The ratio of electro-osmotic to electrophoretic particle fluxes is

\begin{equation}\label{eq:rat1}
\frac{J_{z,eo}}{J_{z,ep}}
=-\frac{3a'\epsilon \psi_0}{2 \xi e}A,
\end{equation}

\noindent where the particle is assumed to be a sphere of hydrodynamic radius $a'$. 
For water at 293~K, this ratio is $-0.10$~$A$, when calculated using values of $a'=100$~nm, $\psi_0=-75$~mV and $\xi = -500$. This calculation demonstrates that both transport mechanisms are important in the present experiments, while suggesting strong dependence on surface charges and potentials, which are difficult to experimentally determine. The value $\psi_0=-75$~mV is equivalent to surface charge density of $0.3e$~nm$^{-2}$ in 0.1~M KCl, consistent with literature values for track-etched polymers \cite{522,535,671}. 
Note that 

\begin{equation}\label{eq:rat1a}
J_{z,ep}+J_{z,eo}=CE_z\left(\frac{\xi e}{6\pi\eta a'}-\frac{\epsilon \psi_0}{4\pi\eta}A\right),
\end{equation}

\noindent so both contributions are linearly dependent on particle concentration and potential difference. 

A corresponding comparison of electro-osmotic and electrophoretic contributions to ionic current carried by ions \cite{786,517} reveals that electrophoresis dominates in the experimental regime considered here. If the second (electrophoretic) term dominates in Eq.~\ref{eq:tripartite}, pore size can be estimated from ionic current measurements by treating the pore as an isotropic, homogeneous conical conductor with bulk conductance $\rho$ equal to that of the electrolyte (see Eqs.~\ref{eq:Maxwell} and \ref{eq:Maxwellcon} below). This approach has been widely used \cite{759,769,760,752,764,778}, including with conical pores \cite{660,530,503,505}.


Current rectification has been observed using resizable nanopores, and should be studied further, although the polarity of applied potential was not varied during the present experiments. Studies have shown \cite{535,639,522,671,784,785} that rectification is determined by asymmetric geometry, non-uniform surface charge distributions and ionic concentration gradients. In previous work, the current anisotropy observed using elastomeric nanopores \cite{660} was at most 25\% for a transmembrane potential of $\pm$200~mV.  

\subsection{Translocation Event Size}
\label{Size}

When a particle travels through a nanopore at constant applied potential, a momentary change in pore resistance is observed - a resistance `pulse' caused by a translocation event. The characteristics of this pulse yield information on the size, shape and nature of the particle. Qualitatively, it is easy to understand that an insulating polystyrene sphere will `block' the pore, increasing its resistance, as it passes through the membrane; that a smaller sphere will produce a smaller increase in resistance; and that the resistance increase is greatest when the particle passes through the narrowest part of the pore. Nearly forty years ago, deBlois and Bean \cite{760} used classical electrostatics to derive quantitative expressions for pulse magnitudes that have been widely applied in recent studies. In this section, approaches detailed by these researchers and others are recounted, and in some cases extended to incorporate a linear conical geometry which is relevant in the present work. It is of interest to compare the various models using experimental data in order to probe the functional relation between particle size, pore size and pore geometry. 

DeBlois and Bean's approaches are approximations, each of which assumes that the particle is spherical and that the surfaces of the membrane, pore and translocating sphere are uncharged. The full solution of Laplace's equation, which is required in order to solve the classical electrostatic problem, is not even available for the simple geometry of an insulating sphere within a conducting tube (Fig.~\ref{FTube}). Moreover, a real system introduces additional complications such as non-cylindrical pore geometry, solution and surface chemistry effects and aspherical particles. 

\subsubsection{Simple Electrostatic Approaches\label{Size1}}

Recent studies of spherical proteins \cite{759, 764} and other particles \cite{764, 767, 768, 752, 762} have based their interpretation of translocation pulses on the simplest of the approaches presented in \cite{760}, which assumes that the sphere radius $a'$ is small in comparison with the cylindrical pore radius $a$. For the geometry defined in Fig.~\ref{FTube}, we initially consider a pore which is cylindrical ($a=b=a_0$) and very long ($l\gg a$). In this case, the electrical resistance of a pore filled with electrolyte of resistivity $\rho$ is  

\begin{equation}\label{eq:Maxwell}
R_{cyl1}=\frac{\rho l}{\pi a_0^2}.
\end{equation}

\noindent Maxwell's expression for the resistivity of a solution containing insulating spheres $\rho_{eff}$ with volume fraction $f$ is  

\begin{equation}\label{eq:rhoeff}
\rho_{eff}=\rho\left(1+\frac{3f}{2}+...\right),
\end{equation}

\noindent where $f$ is just the ratio of the single particle volume to the pore volume, 

\begin{equation}\label{eq:volfrac}
f_{cyl}=\frac{4a'^3}{3a_0^2l}.
\end{equation}

\noindent The absolute change in resistance during a translocation event, or the amplitude of the resistance pulse, is

\begin{equation}\label{eq:sizeratio1cyl}
\Delta R_{cyl} = R_{cyl2}-R_{cyl1}= \frac{2\rho a'^3}{\pi a_0^4},
\end{equation}

\noindent where $R_{cyl2}$ is the equivalent resistance when the pore contains one spherical particle. Dividing Eq.~\ref{eq:sizeratio1cyl} by Eq.~\ref{eq:Maxwell}, the fractional change in resistance is

\begin{equation}\label{eq:sizeratio1cylB}
\frac{\Delta R_{cyl}}{R_{cyl1}} = \frac{2 a'^3}{a_0^2l}.
\end{equation}

\noindent For spheres of constant size, the absolute resistance change varies as $a_0^{-4}$ and the fractional resistance change varies as $a_0^{-2}$. Therefore, current pulses are most clearly observed when the pore size matches particle size as closely as possible.

The same approach can be extended to different pore and particle geometries, while still assuming that $a_0\gg a'$. 
If $l\gg a_0$ does not hold, end effects are significant, in which case the solution given in Eqs.~\ref{eq:Maxwell} to \ref{eq:sizeratio1cylB} is simply modified by replacing the pore length $l$ with the approximate factor $l+1.6a_0$ \cite{760}. For the present work, we explicitly consider a linear conical pore with end radii $a$ and $b$, in which case the pore resistance is

\begin{equation}\label{eq:Maxwellcon}
R_{con1}=\frac{\rho l}{\pi ab},
\end{equation}

\noindent the volume fraction is

\begin{equation}\label{eq:volfraccon}
f_{con}=\frac{4a'^3}{\left(a^2+ab+b^2\right)l},
\end{equation}

\noindent and

\begin{equation}\label{eq:sizeratio1con}
\Delta R_{con} =\frac{6\rho a'^3}{\pi ab \left(a^2+ab+b^2\right)}.
\end{equation}

\noindent Equations~\ref{eq:Maxwellcon} to \ref{eq:sizeratio1con} are derived using volume integrals over a conical pore, and apply when $a$, $b$ and $a'$ are much smaller than $l$. For a truncated cone in which end effects are significant, note that the pore resistance is simply the series sum of the contributions from the ends and central part of the pore. By analogy with the approximation for the cylindrical pore, we can replace $l$ with $l+0.8\left(a+b\right)$ in Eqs.~\ref{eq:Maxwellcon} and \ref{eq:volfraccon}. We obtain a conical version of Eq.~\ref{eq:sizeratio1cylB} by dividing Eq.~\ref{eq:sizeratio1con} by Eq.~\ref{eq:Maxwellcon}, 

\begin{equation}\label{eq:sizeratio1conB}
\frac{\Delta R_{con}}{R_{con1}}=\frac{6 a'^3}{l \left(a^2+ab+b^2\right)}.
\end{equation}
 
\subsubsection{Other Approaches}

Other methods may be more applicable in the large-sphere cylindrical-pore limit. An approach presented in \cite{760} and attributed to Gregg and Steidley \cite{788} has been used in some experiments with latex spheres \cite{768}. In this method, pore resistance is calculated by integrating the annular section surrounding a spherical particle over the length of the pore. This approach does not require that the particle is significantly smaller than the pore size, but necessarily underestimates the pore or blockage resistance, because a uniform current distribution is assumed, whereas any nonuniformity creates a larger resistance. The absolute change in pore resistance is

\begin{equation}\label{eq:BigSphere}
\Delta R_{cyl} =\frac{2\rho}{\pi a_0}\left(\frac{\sin^{-1}\left(\frac{a'}{a_0}\right)}{\left(1-\left(\frac{a'}{a_0}\right)^2\right)^\frac{1}{2}}-\frac{a'}{a_0}\right).
\end{equation}

\noindent In the limit $a_0\gg a'$, this result differs from the previous small sphere result (Eq.~\ref{eq:sizeratio1cyl}) by a factor of two-thirds. However, Eq.~\ref{eq:BigSphere} is asymptotically valid when the sphere size approaches the pore size \cite{760}. 

DeBlois and Bean more rigorously approximated the problem using a different method, in which the calculated electric field bulges around the insulating sphere contained within the pore. Deflection of field lines is minimal, especially for a small sphere, allowing calculation of an upper limit for $\Delta R_{cyl}$. This method applies for any sphere size when $l\gg a_0$, and gives a more rigorous result than Eq.~\ref{eq:BigSphere} in the regime $a'\ll a_0$, asymptotically approaching the result from Eq.~\ref{eq:sizeratio1cyl} in the small sphere limit. The result may be most appropriate for the intermediate region in which the sphere is smaller than, but of comparable dimensions to, the pore size. It can be approximated using the series expansion \cite{760}  

\begin{equation}\label{eq:BigSphere2}
\Delta R_{cyl} =\frac{2\rho a'^3}{\pi a_0^4}\left(1+1.26\frac{a'^3}{a_0^3}+1.1\frac{a'^6}{a_0^6}\right).
\end{equation}

Zimmermann and Jelsch \cite{681} considered resistance pulses generated by cells, and therefore modelled aspherical particles of finite conductivity. Using their approach, the fractional volume change for a single particle in a cylinder is

\begin{equation}\label{eq:JZ}
\frac{\Delta R_{cyl}}{R_{cyl1}} =f_sf_c\frac{4a'^3}{3a_0^2 l},
\end{equation}

\noindent where $f_s$ and $f_c$ are shape and charge factors respectively. This result reduces to Eq.~\ref{eq:sizeratio1cylB} for a perfectly insulating spherical particle, when $f_s=\frac{3}{2}$ and $f_c=1$. The value $f_s=1$ applies to an extended cylindrical particle, similar to an elongated polymer. For a perfectly conducting sphere, the charge factor is $f_c=-\left(f_s-1\right)^{-1}$, so there is a continuum of possible fractional resistance amplitudes which includes zero. This analysis gives some simple insight into the dependence of translocation current peaks on charge distributions, and the associated solution chemistry. Polystyrene is an effective insulator, but surface charges, polarization charges and associated screening charges are present in any nanofluidic system. For example, screening can reduce the value of $\xi$ by more than two orders of magnitude for carboxylated polystyrene spheres \cite{780}. 

\section{Experimental\label{MM}}

Experiments were carried out using a 0.1~M KCl solution, prepared using deionised water (18.2~M$\Omega$) and buffered at pH~8.0 using 0.01~M tris base (p$K_a$ 8.1). Small amounts of Triton X-100 and EGTA were added to aid wetting and as a chelating agent respectively. Reagents were purchased from BDH and Sigma. Particle concentrations are calculated relative to `undiluted' (as-received) aqueous solutions of carboxylated polystyrene spheres (1\%~wt/v, Corpuscular). Nominal particle diameters of 200~nm and 800~nm were used.

\begin{figure}
\begin{center}
\subfigure[]{\label{Cruc}\includegraphics[width=4.5cm]{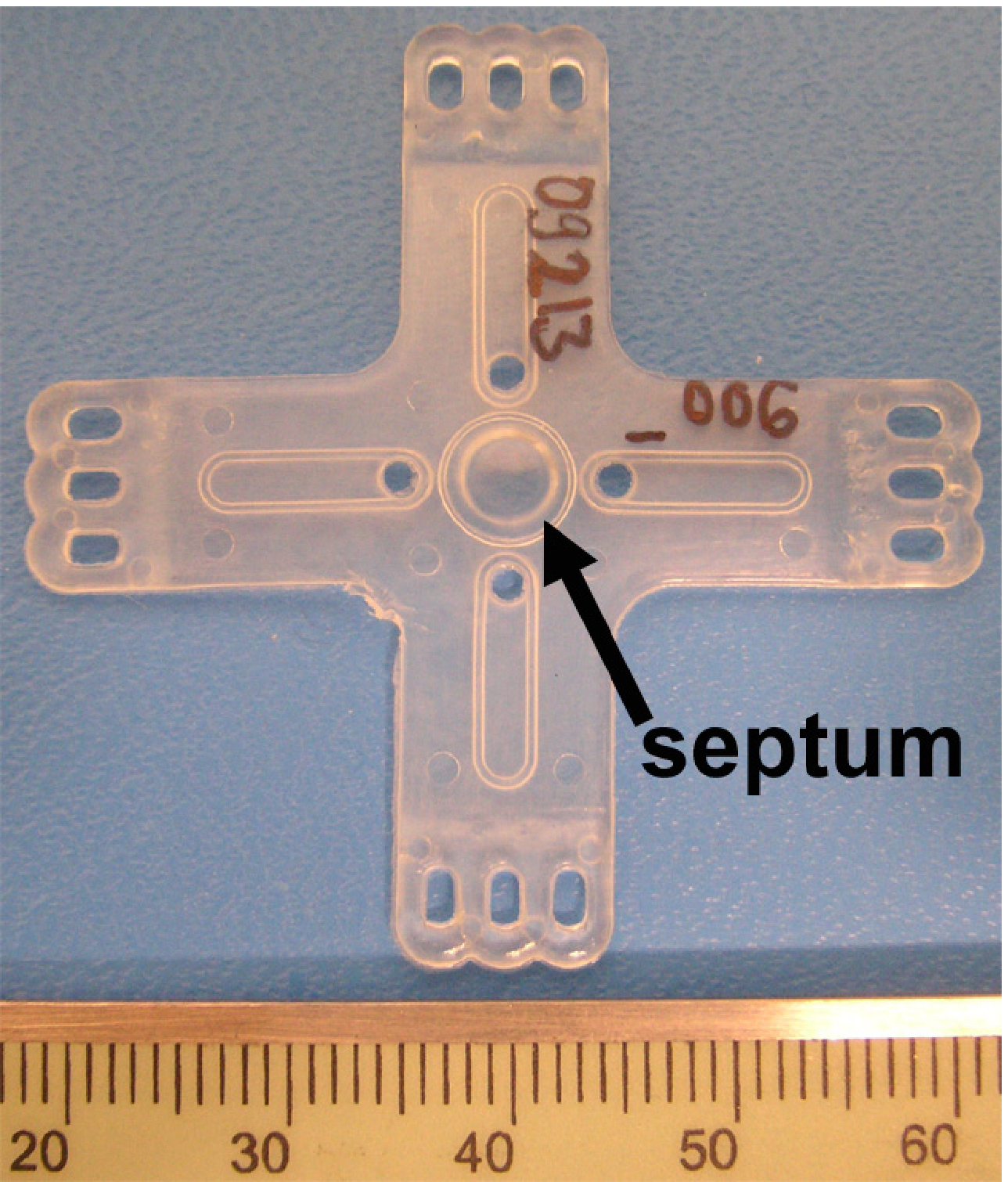}}
\subfigure[]{\label{FluidCell}\includegraphics[width=5cm]{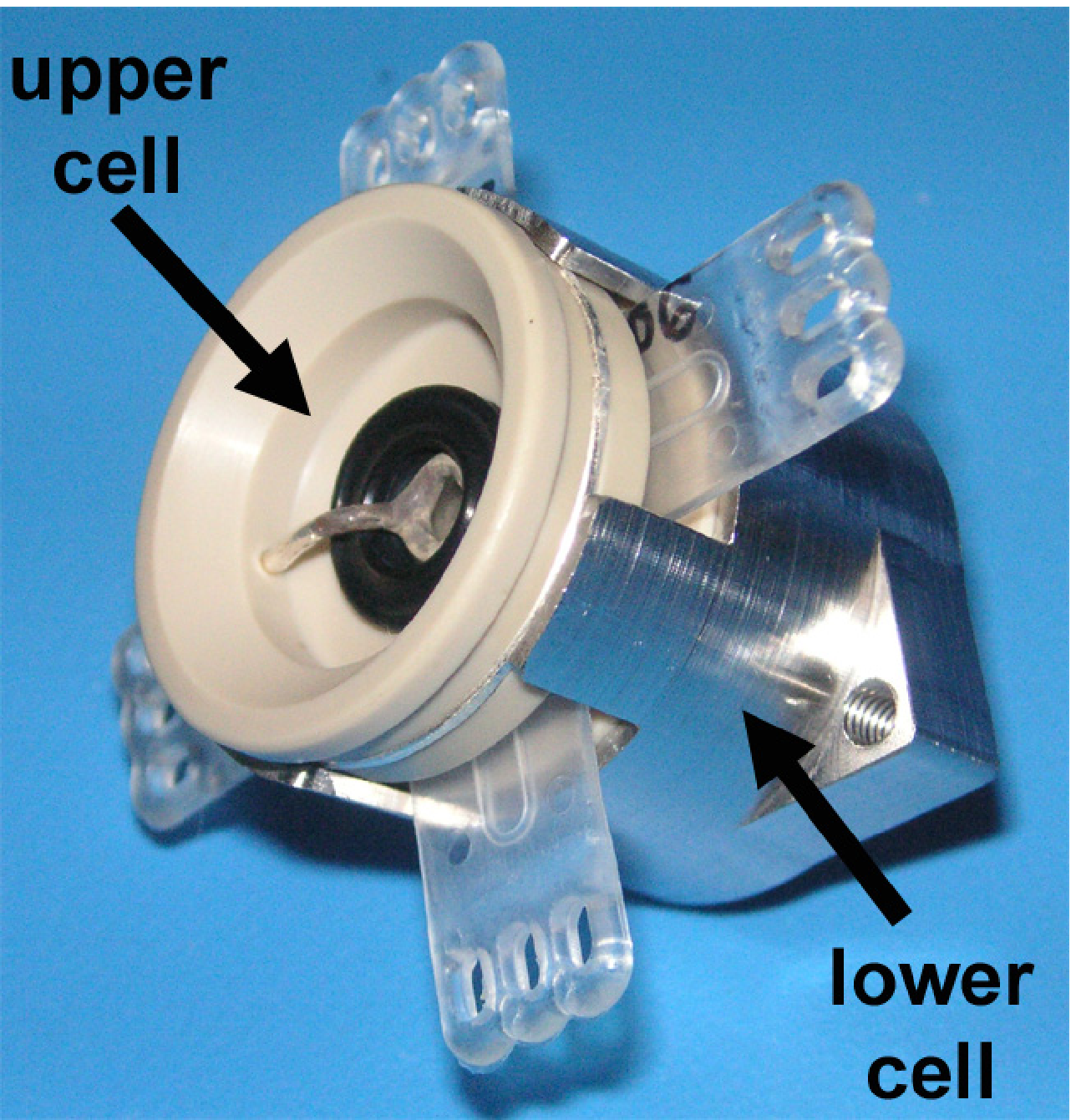}}
\subfigure[]{\label{qNano}\includegraphics[width=5.5cm]{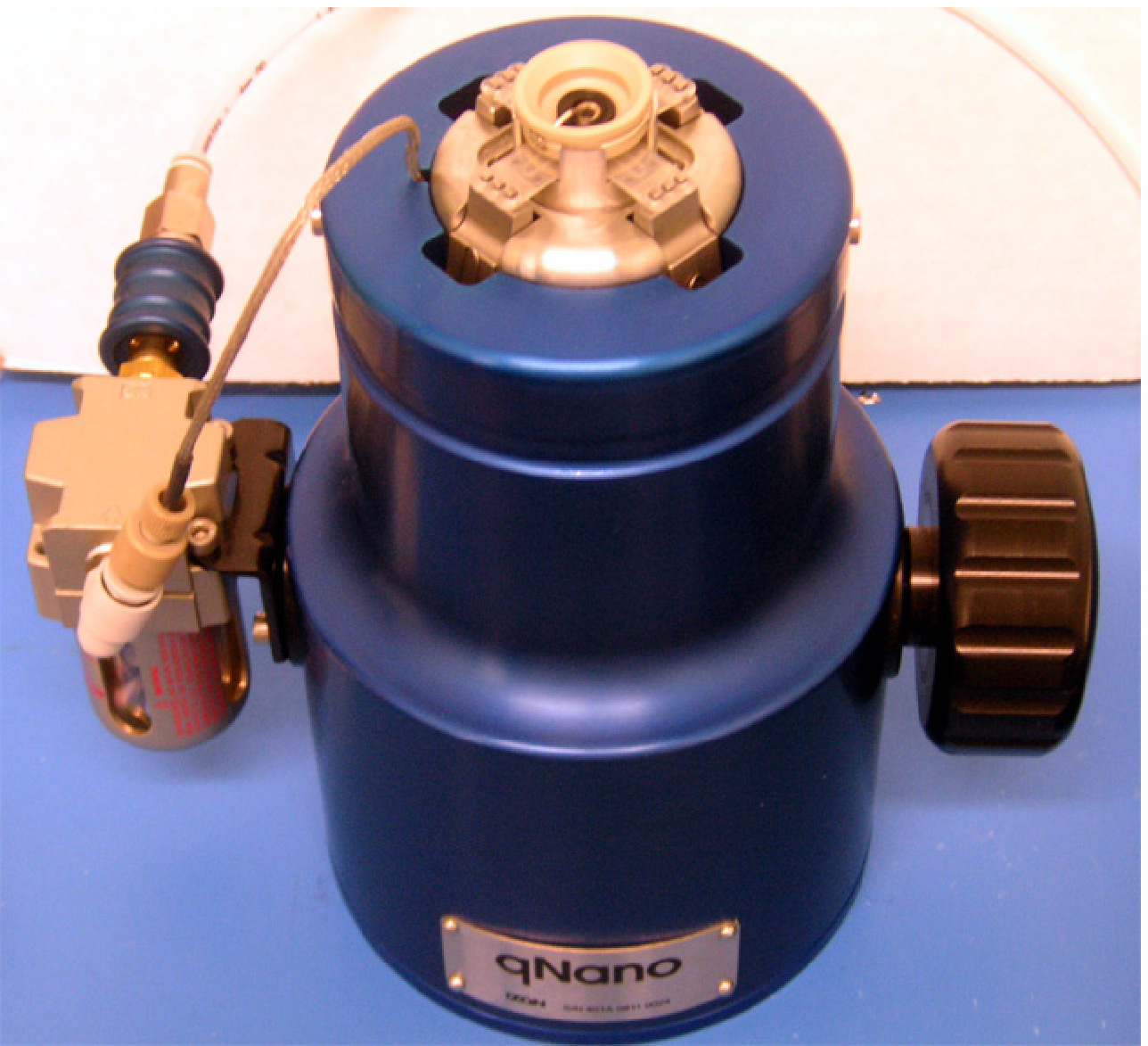}}
\end{center}
\caption{Apparatus used in the experiments. The circular central septum of a TPU cruciform (Fig.~\ref{Cruc}, with a millimetre scale), where nanopores are located, is 200~$\mu$m thick. The remainder of the cruciform is predominantly $\sim$0.8~mm thick. Cruciforms are placed within a fluid cell (Fig.~\ref{FluidCell}) so that o-rings form a loose seal on the cruciform surface. The lower half of the cell is enclosed, whereas solutions can be added or `flushed' into the open upper surface. The cell is placed on the top of the Izon qNano (Fig.~\ref{qNano}) so that teeth fit in the holes at the ends of the cruciform legs. These teeth are actuated by turning the dial on the side of the qNano.}
\label{ExptSetup}       
\end{figure}

Three resizable pore specimens were used. Each pore is located at the centre of a thermoplastic polyurethane (TPU) `cruciform' \cite{454,660,738} (Fig.~\ref{Cruc}). They are fabricated by puncturing the 200~$\mu$m-thick central septum using a specially etched, sharpened tungsten tip. The geometry of each pore is roughly a truncated circular cone, tapering from a large hole on the `cis' surface to a smaller aperture near the `trans' surface. For pores manufactured under the same conditions as the specimens used here, SEM images have revealed cis surface openings of characteristic radius 15~$\mu$m \cite{787, 660, 738}. AFM imaging \cite{660, 738} suggests that the narrowest part of the pore, where the electronic sensing is most sensitive, is within $\sim$1~$\mu$m of the trans surface opening, rather than on the surface itself.

Pores are actuated by extending the distal ends of the cruciform legs, causing the central membrane to stretch radially, with azimuthal symmetry \cite{660}. Each specimen undergoes stress-softening prior to use, so actuation is largely reversible and reproducible outside the near-pore region, which is catastrophically overstretched during pore formation \cite{738}. Following actuation, the cis side pore radius, and the thickness of the membrane, can be estimated by assuming that the material is Hookean, elastic and incompressible \cite{771}. Improved assessment of the relationship between macroscopic strain and deformation local to the pore, including deviations from a linear stress-strain profile, is the subject of ongoing work.  

\begin{figure}
\begin{center}
\subfigure[]{\label{rawA}\includegraphics[width=8.5cm]{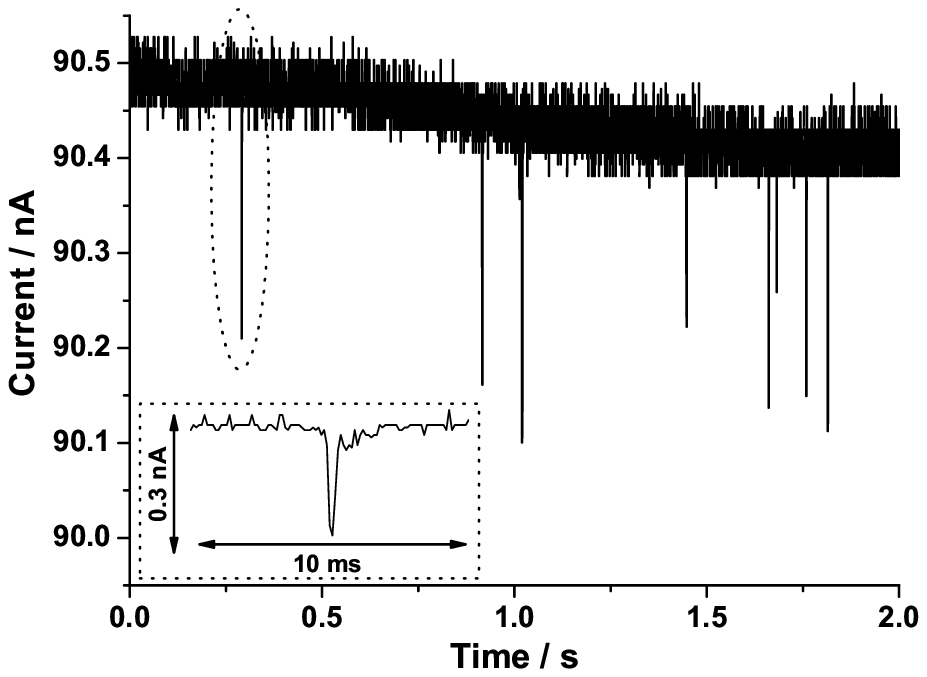}}
\subfigure[]{\label{rawAa}\includegraphics[width=8.5cm]{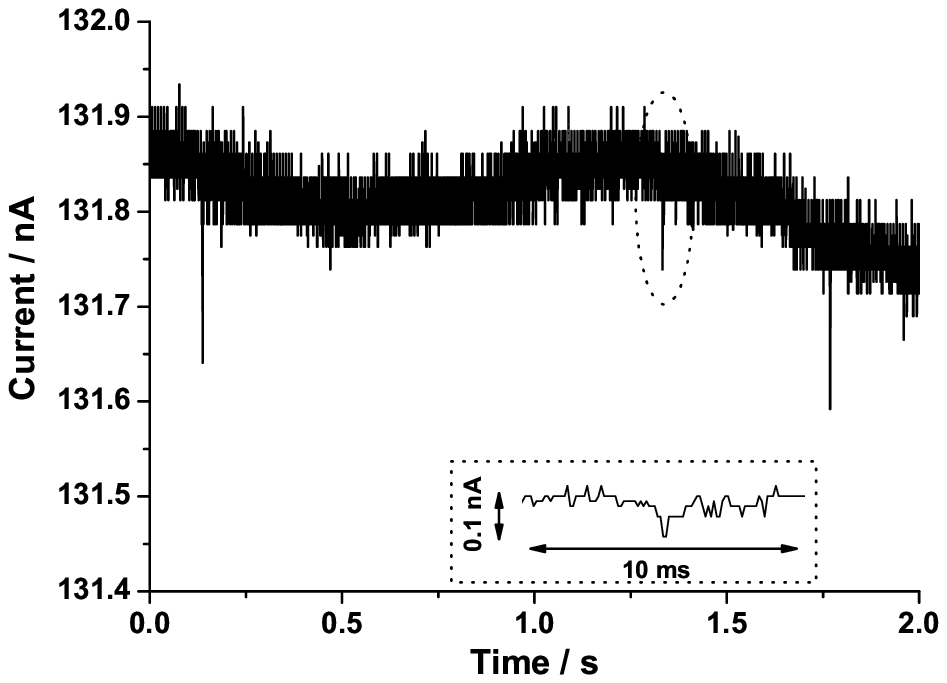}}
\subfigure[]{\label{rawB}\includegraphics[width=5.5cm]{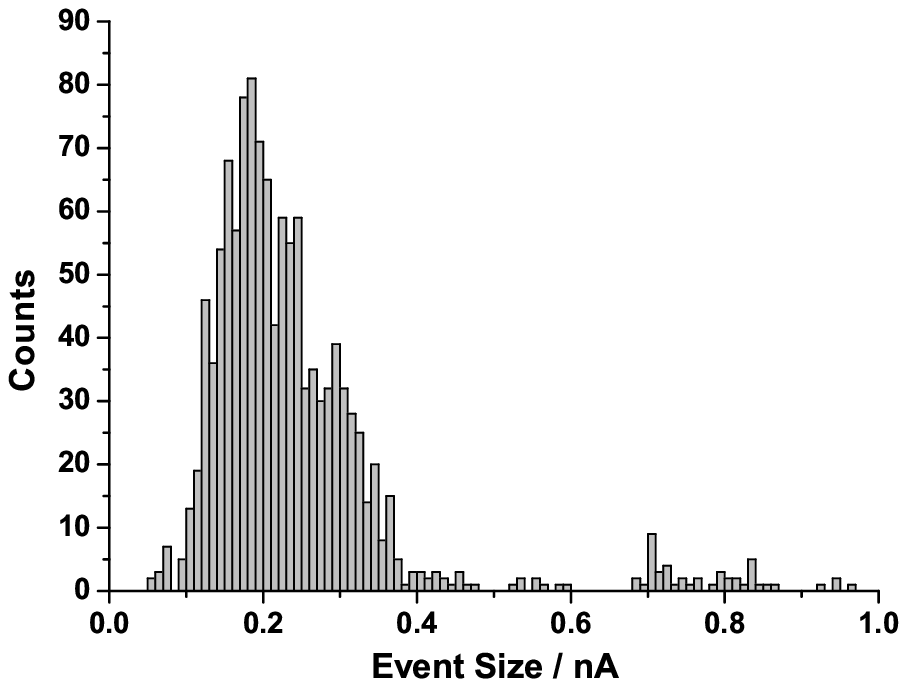}}
\subfigure[]{\label{rawC}\includegraphics[width=5.5cm]{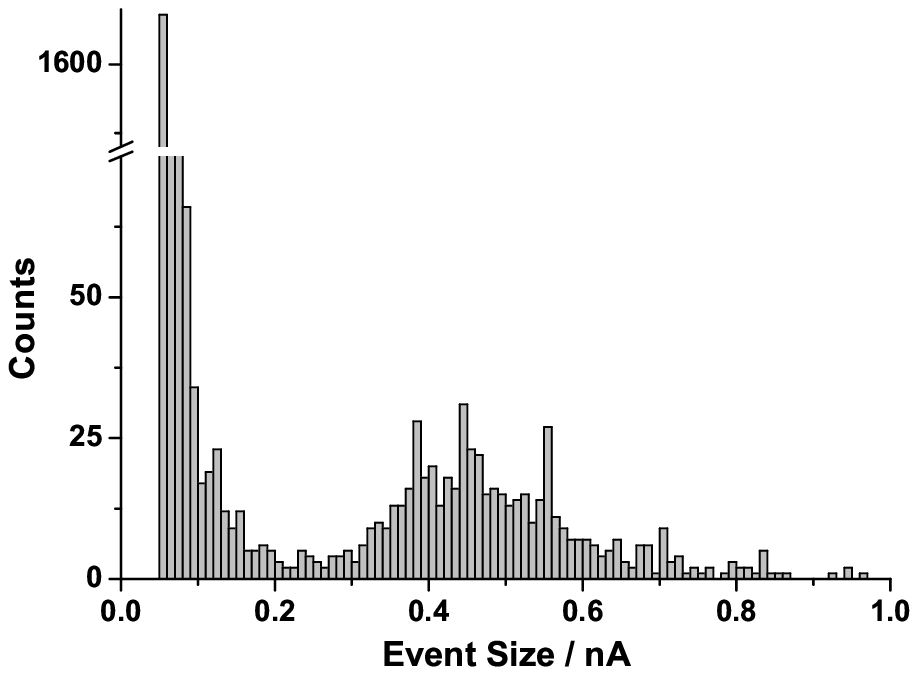}}
\subfigure[]{\label{rawD}\includegraphics[width=5.5cm]{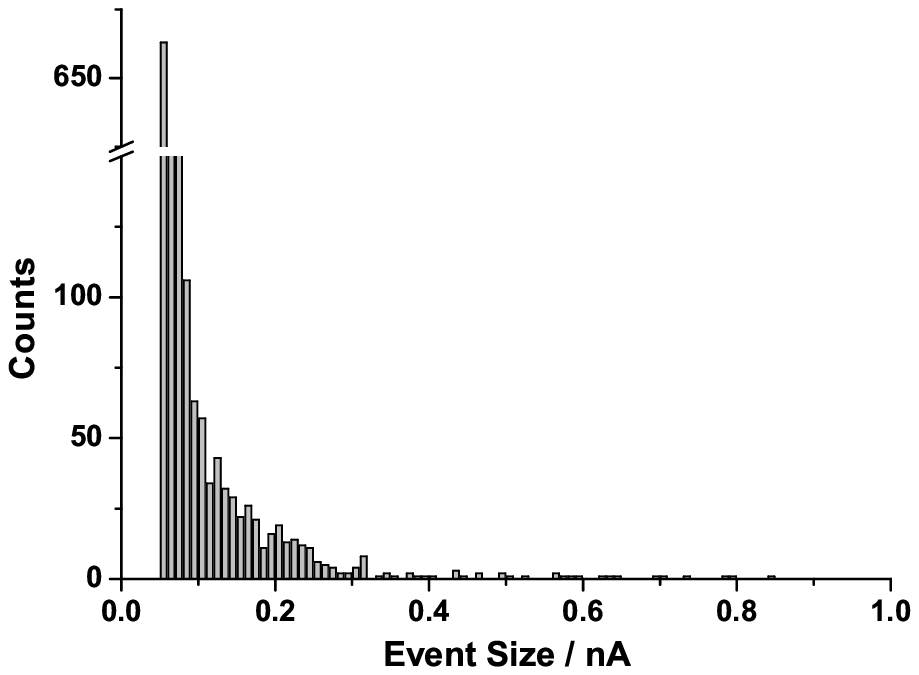}}
\end{center}
\caption{Examples of raw data and event histograms. Figures~\ref{rawA} and \ref{rawAa} show excerpts from typical raw current records using cruciform B (see Table~\ref{Tab:Specimens}). Typical events, indicated by a dashed ring on the main trace, are expanded in the inset traces. In Fig.~\ref{rawA}, from series 1, events were caused by 800~nm spheres and the applied potential was 0.42~V; Fig.~\ref{rawAa} is from series 2, using 200~nm spheres and the potential was 0.44~V. Figures~\ref{rawB}, \ref{rawC} and \ref{rawD} are event size distribution histograms from, respectively, cruciform C series 1 at 0.40~V (1 minute duration), cruciform B series 1 at 0.50~V and cruciform B series 1 at 0.30~V (both 5 minutes duration). Event sizes were characterised relative to the smoothed background current in the vicinity of each event \cite{738} and collated into 10~pA bins.}
\label{Raw}       
\end{figure}

A qNano system (Izon Science, Christchurch, Fig.~\ref{ExptSetup}) was used to actuate the cruciforms and carry out measurements. The central part of each cruciform was wet by electrolyte using a fluid cell which allowed loading of fluid into a reservoir below the specimen, while solutions containing particles could be pipetted on to the open top half of the cell. Vernier callipers were used to measure the distance between the outer rims of holes on opposite cruciform legs. The cruciform `stretch' (see Table~\ref{Tab:Specimens}) is defined as the difference between this measurement in the stretched and relaxed states, where the latter value is 42~mm. Current was measured and voltage applied using Ag/AgCl electrodes in contact with each half of the fluid cell. The current signal (e.g. Figs.~\ref{rawA} and \ref{rawAa}) was digitized at 10~kHz using the qNano's built-in electronics, while software customised for the qNano (Izon Instrument Control System v2.1) was used to set the transmembrane potential, identify events from the current signal and collate the recorded data into histograms.  

Typical records of raw data and examples of event size distributions are shown in Fig.~\ref{Raw}. The current and translocation traces are comparable to other work, using streptavadin for example \cite{768}. The noise threshold for event size, below which events are ignored, is set at 0.05~nA, consistent with the characteristic peak-to-peak baseline signal noise. This current threshold is higher than in other nanopore studies in which molecular-scale particles have been detected. We emphasize that these are initial experiments using comparatively large pores with a portable, desktop system under ambient laboratory conditions. In control experiments with no particles present in the electrolyte, a maximum of 1~event~s$^{-1}$ was identified by the qNano system. All such events had magnitude between 0.05 and 0.06~nA. When nanospheres are introduced to one half of the fluid cell, increased noise is generated by particle activity, such as partial blockages of the pore and interactions of spheres with each other and the pore walls. Fig.~\ref{Raw} demonstrates that events near the noise threshold may be more frequent than a histogram peak caused by translocations at greater event magnitude, or less frequent, as in Figs.~\ref{rawB} and \ref{rawC} respectively. This translocation peak can be dominated by noise (as in Fig.~\ref{rawD}), especially for smaller particles, due to the dependence of event size on particle volume in the simple model (Section~\ref{Size}). In such cases, the tally of events is taken to represent particle `activity', rather than just translocations; this approach is consistent with similar work elsewhere \cite{765}.

The two key parameters extracted from the raw data for the Results are (i) the size of translocation blockage signals and (ii) the frequency of events. The histograms demonstrate that it is not useful to calculate the translocation event magnitude using a mean of all data. It is difficult to separate translocation events from noise events when the tails of the distributions overlap. Therefore, the characteristic size of a translocation current blockage is defined here as the modal peak once the peak at the noise level is discarded. When there is no discernable translocation peak (e.g. Fig.~\ref{rawD}), the data is usually discarded, but the peak at the noise threshold can be interpreted as an upper bound on the actual translocation blockage magnitude. This approach is comparable to the identification of `clusters' used elsewhere \cite{764}.

In order to ensure consistency in frequency data, all identified events above the 0.05~nA noise limit are included. Therefore, frequency data represents all activity, not just translocations. Use of this approach is supported by the smooth linear relationship between event frequency and concentration observed in the Results. The smoothed baseline current adjacent to each event was also recorded in order to calculate the fractional event size, and for estimation of pore size using the bulk conductivity method.

Event duration can be determined by considering the current signal around the event in detail, as in the inset to Figs.~\ref{rawA} and \ref{rawAa}. This parameter is not analysed in the present work due to inconsistency of the data. As discussed above, it is difficult to discriminate translocations from other events. The beginning and end of an event can also be ill-defined, especially when it is considered that the spheres are passing through a long cone. Further work is required to address these issues, or to set up experiments so that pulse durations are more regular.  

Experiments are grouped in specific `series' for each cruciform, with each series corresponding to a stable cruciform setting at which translocations were observed, as shown in Table~\ref{Tab:Specimens}. Within each series, one variable, usually voltage, was altered while a number of separate data files (`runs') were recorded for analysis. For cruciforms A and B (series 1-4), each data point at a particular voltage represents a run of typical duration between 5 and 10 minutes. For cruciform B series 5 and cruciform C, each data point represents the average value from several runs (typically four or five), each of typical duration 30~s to 1~minute. Voltage was generally varied in random chronological order to avoid systematic error.

\begin{table}
\caption{Summary of cruciform specimens, experimental `series' and the associated experimental conditions. As described in the main text, concentration is relative to as-received solutions and the cruciform stretch is measured relative to a resting length of 42~mm, which is also used to calculate strain.}
\label{Tab:Specimens}
\begin{center}
\begin{tabular}{ccccccc}\hline\hline
Cruciform&Series&Stretch&Strain&Concentration&Sphere Radius&Voltage Range\\
&&mm&&x $10^{-4}$&nm&mV\\\hline
A&1&10.3&0.25&2&400&0.16-0.30\\
&2&10.4&0.25&2&400&0.16-0.30\\
&3&16.0&0.38&2&400&0.16-0.30\\
&4&19.1&0.45&2&400&0.16-0.30\\\hline
B&1&19.2&0.46&1&400&0.34-0.50\\
&2&19.2&0.46&1&100&0.38-0.46\\
&3&10.0&0.24&2&400&0.12-0.50\\
&4&10.0&0.24&2&100&0.16-0.49\\
&5&9.51&0.23&Varied&400&0.30 only\\\hline
C&1&13.5&0.32&2&400&0.12-0.46\\
&2&13.5&0.32&2&400&0.14-0.44\\\hline
\hline
\end{tabular}
\end{center}
\end{table}

\section{Results} 

\subsection{Dependence of Event Frequency on Concentration and Voltage}

\begin{figure}
\begin{center}
\subfigure[]{\label{FigA}\includegraphics[width=8.5cm]{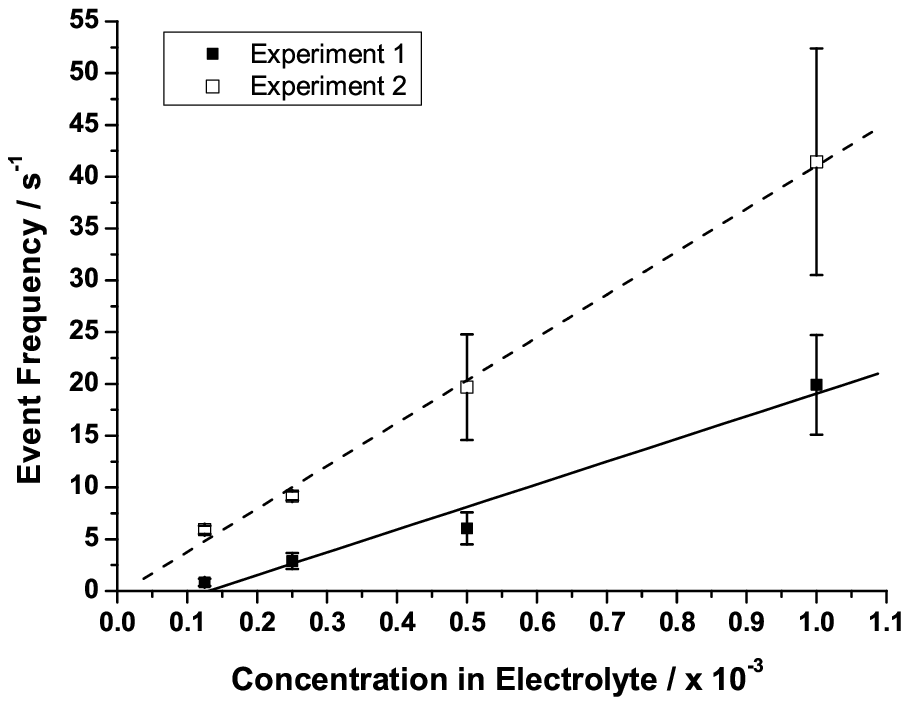}}
\subfigure[]{\label{FigB}\includegraphics[width=8.5cm]{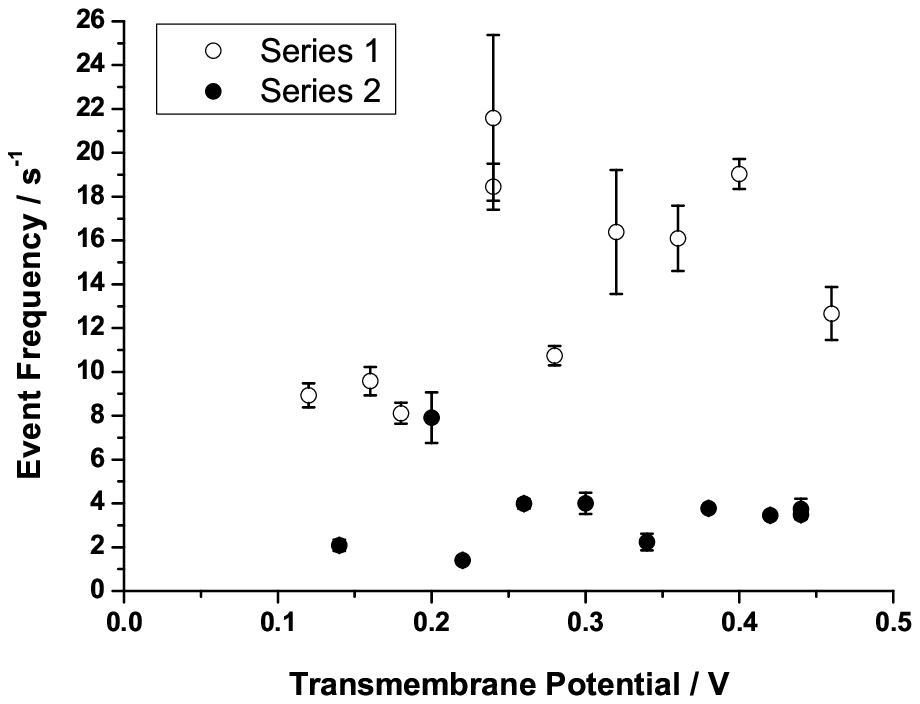}}
\end{center}
\caption{Figure~\ref{FigA} is a plot of event frequency as a function of concentration (relative to as-received colloidal solutions) at 0.30~V applied transmembrane potential (cruciform B, series 5). The two experiments represent data captured on consecutive days. $R^2 = 0.973$ and 0.997 for linear fits to experiments 1 and 2 respectively. Figure~\ref{FigB} shows event frequency plotted over a working range of voltages for two series using cruciform C. $R^2$ values for a linear fit were less than 0.5 for these series, which were recorded at constant particle concentration and pore resistance. Plotted points and error bars represent the mean and standard error in the mean for multiple recordings.}
\label{FigsAB}       
\end{figure}

The data in Fig.~\ref{FigA} 
are consistent with the predicted electro-osmotic and electrophoretic particle flux (Eq.~\ref{eq:rat1a}), as well as previous experiments \cite{738}. It is notable that the trend does not pass through the origin, suggesting that the simple transport model of Section \ref{PFlux} does not extend to low concentrations. Data plotted in Fig.~\ref{FigB}, which are typical of other experiments, suggest a weak positive trend between event frequency and applied voltage, especially if the outlying data with relatively large error bars is discounted. The transport model suggested in Section \ref{PFlux} 
predicts that the event frequency should increase linearly with applied voltage. There is little evidence for such a trend; any linear positive trend would appear to intercept above the origin. In particular, the significant scatter in the data appears inconsistent with the result from Fig.~\ref{FigA}.

There are several possible reasons for the inconsistency between Figs.~\ref{FigA} and \ref{FigB}. In a broad sense, it is most likely that the data reflects differences between the two specimens used. One other likely explanation is that pore morphology changes unpredictably with applied voltage, producing the marked difference in the scatter of the data in the two figures. The polymeric internal pore surface and adjoining electrolyte include regions of aggregated charge that could experience a force in an applied electric field. Parts of the pore surface near the constriction can be ragged and mechanically unconstrained due to the fabrication method. 

Other factors could explain the nonlinear relation or generally poor data consistency observed in Fig.~\ref{FigB}, but not necessarily both. The simplifying assumptions required in order to arrive at Eq.~\ref{eq:rat1a} should not explain the scatter in the data. Relevant issues would include any electric field leakage near the tip of the cone, the suitability of the Space-Charge approach for a conical geometry, and possible variations in particle concentration when the pore diameter approaches the particle size. Three further experimental factors could be significant. Firstly, the number of recorded events represents all activity, rather than just translocations (discussed above). Secondly, apparently spontaneous changes between distinct, stable `modes' are observed when using this device, evidenced by step changes in baseline current and possibly caused by adhesion of particles to the polymer surface. However, switches between such modes are typically identified if they are significant. Thirdly, it is possible that a small pressure difference is introduced across the membrane as it is sealed to the lower half of the fluid cell.  

\subsection{Translocation Resistance Pulse Sizes}

\subsubsection{Voltage Dependence}

\begin{figure}
\begin{center}
\subfigure[]{\label{Fig1}\includegraphics[width=8.5cm]{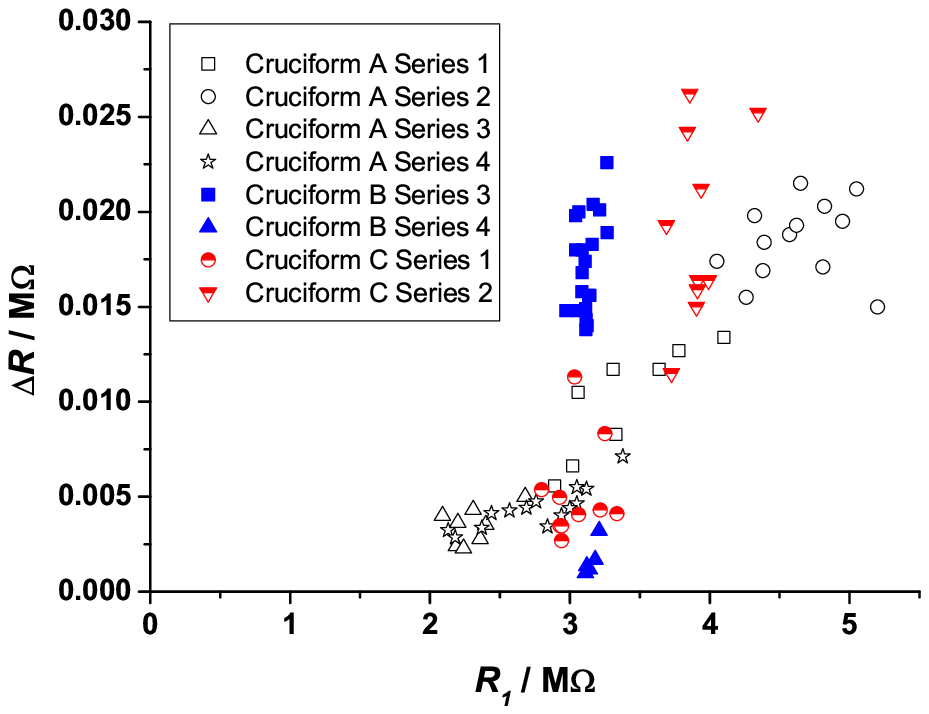}}
\subfigure[]{\label{Fig2b}\includegraphics[width=8.5cm]{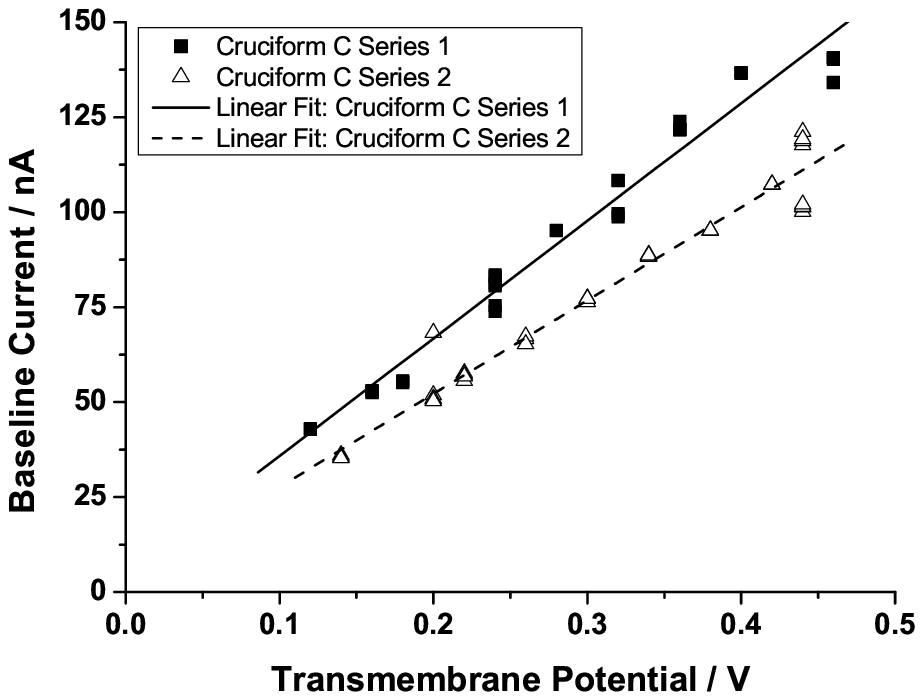}}
\end{center}
\caption{Figure~\ref{Fig1} plots the modal resistance change of events against baseline nanopore resistance. Each data point represents the average of data collected at a particular applied potential. 200~nm particles are used in cruciform B series 4, for which the data represent an upper bound on translocation resistance pulse magnitude (see Section~\ref{MM}); 800~nm particles are used for the other data. In Fig.~\ref{Fig2b}, baseline current is plotted against applied potential for experiments using cruciform C. $R^2=0.97$ for linear fits to the data in both cases.}
\label{Fig1and2b}       
\end{figure}

Figure~\ref{Fig1} shows the modal size of the resistance pulses associated with translocation events. Overall, there is a rough positive trend between the size of a resistance pulse and the baseline resistance. The exact relation is inconsistent between different series due to geometrical variations relating different pores, stretch states and particle sizes. These observations do not depend strongly on the applied transmembrane potential, consistent with the analyses in Section~\ref{Size}. It is apparent in Fig.~\ref{Fig1} that the range of $R_1$ values is limited for any given series. Further, Fig.~\ref{Fig2b} demonstrates that baseline resistance is typically characterised by an Ohmic $I$-$V$ plot over the range of interest. Although rectification was not specifically studied here, the Ohmic characteristic suggests that any current asymmetry is not significant in the interpretation of results.

Experiments using cruciform B demonstrate smaller resistance pulses for smaller particles in two series at the same cruciform strain. This trend is intuitive and qualitatively in line with the theoretical approaches. Quantitatively, there is a factor of $\sim$10 difference between the modal resistance peak for 800~nm particles and the upper bound on that peak for 200~nm spheres, using a cruciform held at the same strain (Fig.~\ref{Fig1}). This is lower than the difference predicted (e.g. Eq.~\ref{eq:sizeratio1cyl}) if the blockade magnitude scales with the sphere volume. Taking into account that the peak for 200~nm spheres is an upper bound, the data therefore suggest that improved resolution is required to accurately study particles of such disparate sizes using a static pore. A key method of improving the signal-to-noise ratio is to match the particle size more closely to the pore size, a process which is directly enabled by resizable nanopore technology and which will be the subject of further work.

\subsubsection{Estimates of Pore Radius}

\begin{figure}
\begin{center}
\includegraphics[width=8.5cm]{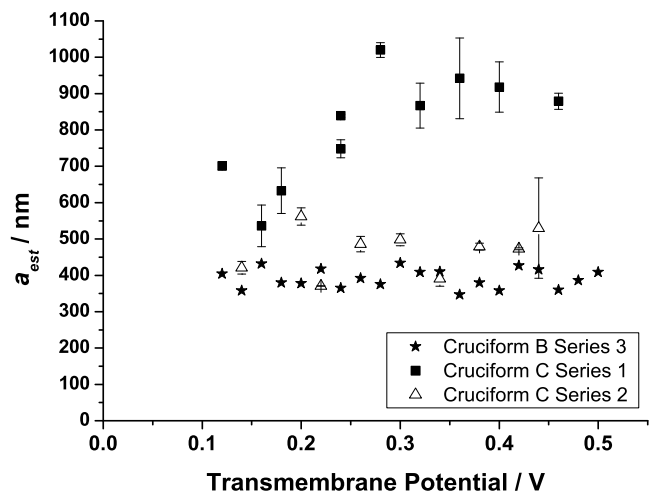}
\end{center}
\caption{Estimated pore radius $a_{est}$, calculated from the fractional event size using Eq.~\ref{eq:sizeratio1cylB}, is plotted as a function of applied potential. Stretch-induced membrane thinning has been accounted for in the calculation, and each data point represents the average of all events recorded at a particular voltage. For the cruciform C series, error bars represent the standard error in the mean of several, relatively short runs at the same potential.}
\label{Fig2a}       
\end{figure}

\begin{table}
\caption{Comparison of the theoretical approaches to calculating $a_{est}$, as applied to two data series obtained under similar experimental conditions (see Table~\ref{Tab:Specimens}). Thinning of the stretched membrane is treated as described in \cite{771}. All of the models assume that $a_{est} \ll l$, and uncertainties are dominated by choice of the model used for calculations, as discussed in the text. Estimates refer to the smaller pore radius in the conical case, and to the narrowest part of the pore when using the event resistance change.}
\label{Tab:a_eff}
\begin{center}
\begin{tabular}{cccc}\hline\hline
Model for $a_{est}$&Assumes $a'\ll a_{est}$?&Cruciform C Series 1&Cruciform C Series 2\\
&&$\mu$m&$\mu$m\\\hline
From Baseline Resistance:&&&\\ 
Cylinder (Eq.~\ref{eq:Maxwell})&Yes&$3.5$&$3.1$\\
Cone (Eq.~\ref{eq:Maxwellcon})&Yes&$0.6$&$0.5$\\\hline
From Event Resistance Change:&&\\ 
Cylinder (Eq.~\ref{eq:sizeratio1cyl})&Yes&$1.6$&$1.2$\\
Cylinder, fractional (Eq.~\ref{eq:sizeratio1cylB})&Yes&$0.7$&$0.4$\\
Cylinder, lower bound (Eq.~\ref{eq:BigSphere})&No&$1.5$&$1.1$\\
Cylinder, upper bound (Eq.~\ref{eq:BigSphere2})&No&$1.6$&$1.2$\\
\hline\hline
\end{tabular}
\end{center}
\end{table}

The estimated pore radius derived from experimental data, denoted $a_{est}$, is heavily dependent on the model used for the calculation. At present, this value should be treated as an indication of effective pore size for comparative purposes, rather than necessarily an accurate determination of pore radius. Typical data showing the calculated estimated radius $a_{est}$ as a function of applied potential are shown in Fig.~\ref{Fig2a}. These data are calculated using the modal fractional event size (Eq.~\ref{eq:sizeratio1cylB}) at each potential. In each case, $a_{est}$ is calculated with the nanopore in a stressed state, with changes of the membrane thickness $l$ estimated using the simple approach described in \cite{771}. This plot further explores the assertion of voltage independence and an effectively Ohmic relationship observed in relation to Fig.~\ref{Fig1and2b}, and consistent with the analyses in Section~\ref{Size}.

For most of the data series in Table~\ref{Tab:Specimens}, results are similar to those plotted in Fig.~\ref{Fig2a} for cruciform B series 3 - consistent over the working range of voltages. For cruciform C series 1, there is an indication of a positive trend in estimated pore size with voltage. Such a trend is atypical of the data for all series, although the degree of scatter is consistent with the other data sets. Comparing the two data series for cruciform C, it is clear that apparent pore dimensions can change between series in which experimental conditions are repeated. The error bars for the cruciform C series indicate that the results for a particular series, applied potential and sphere size are comparable to the scatter over a range of potentials and possible differences between series, as indicated by variations in estimated pore size.

Various methods of calculating estimated pore size $a_{est}$ from current measurements are compared in Table~\ref{Tab:a_eff}. Three models, using absolute resistance data with Eqs.~\ref{eq:sizeratio1cyl}, \ref{eq:BigSphere} and \ref{eq:BigSphere2}, give very similar results. It follows that the assumptions differentiating these approaches are relatively insignificant in the pore-sizing process. These assumptions are (i) $a' \ll a_0$, (ii) the electric field is uniformly distributed for Eq.~\ref{eq:BigSphere}, and (iii) the electric field bulges as required for Eq.~\ref{eq:BigSphere2}. In contrast, the method of dealing with pore geometry (including end effects) significantly varies between these three models and each of the other models. 

DeBlois and Bean's simplest analysis can be applied to both absolute (Eq.~\ref{eq:sizeratio1cyl}) and fractional (Eq.~\ref{eq:sizeratio1cylB}) resistance measurements. The absolute case, which assumes cylindrical pore geometry and does not depend on $l$ or $b$, gives a significantly smaller pore size than the cylindrical result from baseline resistance, because the absolute pulse is dominated by the signal when the sphere is within the narrowest part of the conical pore - at the small pore end. Values calculated using Eqs.~\ref{eq:BigSphere} and \ref{eq:BigSphere2} use a similar geometric approach. Results calculated using the fractional resistive pulse data are two to three times smaller than the absolute resistance data. This is expected, because $a_{est}$~$\propto$~$(R$~/~$\Delta R)^2$ for the fractional data and $a_{est}$~$\propto$~$(1$~/~$\Delta R)^4$ for the absolute data; the measured value of $R$, averaged over the length of the pore, is relatively low, whereas the value of $\Delta R$ is mostly dependent on the narrowest part of the pore. 

As expected, baseline resistance data gives the largest value for the cylindrical case, because that model assumes consistent resistive losses along the pore length. For the conical model, resistive losses are concentrated near the smaller end of the truncated cone. The conical calculation uses estimated values of the larger pore radius $b$, determined using SEM images of typical pores at rest and adjusted to reflect membrane stretching using the approach described in \cite{771}.

Typically, the standard error in measurement of resistance or modal event size is $<5$\%. Uncertainties in the calculated values from Table~\ref{Tab:a_eff} are dominated by geometric factors, most of which have been discussed above in comparing the different methods of modelling. To illustrate this point, we note that the two further conical models (Eq.~\ref{eq:sizeratio1con} and \ref{eq:sizeratio1conB}) give outlying results ($a_{est}<10$~nm) because of their dependence on $b$. The probable explanation for this observation is that the profile of the pore, rather than being a linear truncated cone, is convex (trumpet-shaped) \cite{738}. It follows that the volumetric ratio in Eq.~\ref{eq:volfrac} is an underestimate. If this explanation is correct, the calculation of $a_{est}$ using the conical, baseline resistance method should also be an underestimate; this is entirely reasonable given that the smaller pore radius should accommodate the translocating particles. It should further be noted that the model for membrane thinning and pore resizing \cite{771} uses linear elastic material parameters. The approaches which use the experimental value $\Delta R$ have the disadvantage of introducing greater fractional random error associated with this differential measurement. However, the greatest measurement uncertainty in the present experiments is associated with the larger pore radius $b$, which is typically $\sim 15\pm5$~$\mu$m. Improvement to techniques for imaging these pores is ongoing \cite{787}. Other issues, which are less consistent with observed differences between model calculations, could arise from the method of electronic sampling and from significant charge effects. 

There would be great practical benefits associated with determining pore size and shape using simple current measurements. A reliable method would enable expedient use of the technology, without the need for expensive, time-consuming and potentially destructive methods of specimen characterisation. Of the approaches considered, the most reliable are likely to be those models using absolute pulse data with cylindrical geometry. These models estimate the pore size where the particle is most constricted within the pore, so the value of $a_{est}$ should refer to the characteristic pore size at or near this constriction. These approaches do not require detailed knowledge of the pore profile, or the values of $b$ and $l$, so modelling of stretch effects is not required. Improved control and understanding of pore geometry will be key to developing pore-sizing techniques, while charge effects are likely to become more significant for smaller pores.  

Further development of the analytic current blockade analyses could extend the models for which $a'\ll a_0$ does not hold to a conical pore, in which case the effective solution resistance is not constant down the length of the pore. A range of more realistic, non-conical profiles such as those used by Ram\'{\i}rez et al. \cite{671} could be considered. End effects might be important when a particle is near the most active, narrow end of the nanopore. The usual assumption for neglecting the pore ends is $a\ll l$, so significant effects are likely when the particle is within a few microns of the narrow pore opening. The total population of spheres within the cone of the pore could also be considered. The presence of multiple spheres could significantly impact the effective solution conductivity and current noise characteristics. Finally, surface charge effects should be included for application to future experiments using smaller resizable pores.

\section{Discussion and Conclusions}
\label{sec:3}

This paper has presented initial experimental work using novel resizable nanopore technology. Variation of a number of experimental variables has been demonstrated using theory and experiments. Experimental trends agree with those established using other apparatus. For example, the absolute size of experimental translocation resistance pulses increases with the baseline resistance, meaning that translocation signals are clearest for a particular particle set when the resizable pore radius is as small as possible. The increase in pulse size between signals for 200~nm and 800~nm particles is consistent with volumetric scaling of pulse size. There is little dependence of pulse size on applied voltage, a result consistent with a predominantly Ohmic $I$-$V$ relationship, which was independently verified.
The experimental data also demonstrated some inconsistency, manifested as scatter, reflecting that measurements were carried out using a dynamically variable system.

A method for accurately estimating nanopore size from current measurements alone would be valuable for applications of this technology. For that purpose, several theoretical models were directly compared using experimental data, including a novel approach in which truncated cone geometry was applied to resistance pulse magnitudes. The most accurate models rely on absolute resistance pulse height, emphasizing data recorded when the sphere is at the narrowest part of the pore. Such models do not heavily depend on the pore profile through the entire membrane thickness. Overall, the spread of values calculated using different models demonstrated that internal pore geometry is the most important, least well-understood experimental parameter. 

Several issues relating to experimental configuration and data collection have been directly addressed in this work. It was established that both electrophoretic and electro-osmotic flow were probably significant for particle transport in the experiments described. Key functional relationships presented in the theory section allowed effective handling of this experimental regime. In particular, it was predicted that particle flux should be proportional to concentration and electric field. Experiments supported the former relationship, while the latter was less certain. Discrimination of translocations from noise and other particle activity was addressed by specifically identifying modal translocation peaks in the event size distribution, consistently applying a noise threshold across all experiments, and including all activity in the event frequency statistics.
 

Future experiments will probe the theoretical models and experimental variables more closely, and move towards reliable miniaturization of the technology. Improvement of experimental characterization procedures will provide the most significant advances towards efficient, accurate pore-sizing. Important areas for development include assessment of the internal profile of pores and studies of the charge-based solution chemistry, including determination the surface potential on the interior surface of a pore. As Henriquez et al. \cite{767} have suggested, only particle size and concentration can be determined from nanopore experiments when geometry and chemistry are poorly defined. Ongoing theoretical work relating to blockade sizes should concentrate on incorporating charge effects and pore actuation. Further developments relating to geometrical aspects are likely to provide gains in accuracy which are insignificant when compared with the experimental spread of values arising from simple models. 

\section*{Acknowledgements}

The authors would like to thank Izon Science for providing apparatus and assistance. Drs. Mike Arnold, Murray Jansen and Roger Young of Industrial Research Limited contributed to useful discussions. This work has been partly funded by the Foundation for Research, Science and Technology and The MacDiarmid Institute for Advanced Materials and Nanotechnology.

\end{document}